\documentclass[showpacs,pra,a4paper,floatfix,superscriptaddress,twocolumn]{revtex4}
\usepackage{graphicx}
\usepackage{bm,epsfig}
\usepackage[dvips]{pstcol}
\usepackage{amsmath}
\usepackage{amssymb}

\begin{document}

\title{Gain-assisted superluminal light propagation through a Bose-Einstein condensate cavity system}

\author{S. Hamide \surname{Kazemi}}
\email{s.h.kazemi@znu.ac.ir}
\affiliation{Department of Physics, University of Zanjan, University Blvd., 45371-38791, Zanjan, Iran}

\author{Saeed \surname{Ghanbari}}
\email{sghanbari@znu.ac.ir}
\affiliation{Department of Physics, University of Zanjan, University Blvd., 45371-38791, Zanjan, Iran}

\author{Mohammad \surname{Mahmoudi}}
\email{mahmoudi@znu.ac.ir}
\affiliation{Department of Physics, University of Zanjan, University Blvd., 45371-38791, Zanjan, Iran}
\date{\today}

\begin{abstract}
The propagation of a probe laser field in a cavity optomechanical system with a Bose-Einstein condensate is studied. The transmission properties of the system are investigated and it is shown that the group velocity of the probe pulse field can be controlled by Rabi frequency of the pump laser field. The effect of the decay rate of the cavity photons on the group velocity is studied and it is demonstrated that for small values of the decay rates, the light propagation switches from subluminal to superluminal just by changing the Rabi frequency of the pump field. Then, the gain-assisted superluminal light propagation due to the cross-Kerr nonlinearity is established in cavity optomechanical system with a Bose-Einstein condensate. Such behavior can not appear in the pump-probe two-level atomic systems in the normal phase. We also find that the amplification is achieved without inversion in the population of the quantum energy levels.
\end{abstract}
\maketitle

\section{Introduction}\label{intro}
 Quantum coherence has a major role in  controlling the group velocity of light pulses in a dispersive medium. It is well known that, the group velocity of light pulses can be reduced to few meters per second known as slow light. It can also exceed the propagation velocity of light in vacuum and even can become negative, leading to the superluminal light propagation. The superluminal light propagation does not violate the special relativity theory of Einstein. It is wroth noting that, information can not propagate with a velocity more than that of light in vacuum~\cite{Chiao}, however part of information located around the peak of the pulse can propagate  with the group velocity in superluminal region. Various mechanisms have been introduced for controlling the group velocity of light pulses through the atomic systems in normal phase. The gain-assisted superluminal light propagation has been  observed in gaseous systems~\cite{Wang}. It has been also theoretically reported in coupled optical resonators \cite {chang} and in a duplicated two-level system~\cite{duplicated}.  Slow and fast light propagation in solids at room temperature have also been realized based on the process of coherent population oscillations~\cite{Big1,Big2} and by using stimulated Brillouin scattering~\cite{The}. The absorption-free superluminal light propagation has been introduced in a closed-loop atomic system beyond multi-photon resonance condition~\cite{mahmoudi1} and in a three-level pump-probe system~\cite{mahmoudi2}. More recently, the gain-assisted superluminal pulse propagation via four-wave mixing in superconducting phase quantum circuits has been reported~\cite{Amini}.

The optical properties of a probe field in a quantum system, containing a Bose-Einstein condensate, have been investigated. For a many body system consisting of interacting Bosons, below a critical temperature, atoms start occupying the lowest energy level of the many body quantum system having the same quantum state, known as Bose-Einstein condensate (BEC)~\cite{Pethick}. Since the experimental realisation of the atomic Bose-Einstein condensation in 1995~\cite{Anderson,Davis}, which had been predicted by Einstein in 1925, the study of ultracold Bose and Fermi gases has become the most active field of research in atomic physics~\cite{Leggett,Pethick}.
Recently, it has been shown that a collective density excitation of a BEC inside the cavity can serve as a mechanical oscillator coupled to the cavity field, known as a cavity optomechanical system~\cite{Brennecke}. Similar to electromagnetically induced transparency, the optomechanically induced transparency was indroduced by Weis et. al. ~\cite{Weis}. The optical cross-Kerr nonlinear response in a BEC has been investigated and has been proposed as a scheme for all-optical Kerr switch based on the coupled BEC cavity system~\cite{Chen}. Moreover, it was shown that slow light can easily be realized in this system~\cite{Chen2}. Also, the bistable behavior of the photon number in an optical cavity, filled with a BEC, has been reported~\cite{Zubairy}. More recently, fast and slow light propagation in a double-ended cavity with a moving nanomechanical mirror~\cite{tarhan} and in a hybrid optomechanical system, containing a two-level atom~\cite{akram} have been reported.

In this paper, we investigate the effect of the pump field, damping rate of the atomic excited state and the decay rate of the cavity photon on the propagation of the probe field through the BEC-cavity system. It is demonstrated that, the group velocity of the probe field can be controlled by the intensity of the pump field so that for special selected parameters the gain-assisted superluminal light propagation is established due to the cross-Kerr nonlinearity. Note that, the induced-gain is achieved without any inversion in the population of the quantum energy levels.

\section{Model and Equations}\label{Model}
Our system is shown in figure 1, where a cigar-shaped Bose-Einstein condensate of $N$ $\rm{{}^{87}}$Rb atoms is coupled  to the field of an optical ultrahigh-finesse Fabry-Perot cavity.  As the axis of the cigar-shaped BEC and the cavity axis are parallel, the system can be considered one-dimensional.  Two external laser fields, including a pump-laser with central frequency $\omega_{pu}$ and a weak probe-laser with frequency $\omega_{pr}$  are added along the cavity axis.  In the large detuning limit, and after adiabatic elimination of the excited states of the atoms, the effective Hamiltonian of the coupled BEC-cavity system can be written as~\cite{Masch,Horak},
\begin{eqnarray}
\hat{H}_{eff}\ &=& \int \hat{\Psi}^{\dagger}({x}) \{\frac{-\hbar^{2}}{2 m} \frac{d^{2}}{dx^{2}}\\
&+& V_{ext}({x})+\hbar U_{0} \cos^{2}(kx) \hat{a}^{\dagger} \hat{a} \}\hat{\Psi}({x})dx \nonumber\\
&+&\hat{H}_{A-A} + \hbar \Delta \hat{a}^{\dagger} \hat{a}+i \hbar E_{pu} (\hat{a}^{\dagger}- \hat{a})\nonumber \\
&+& i \hbar E_{pr} ( \hat{a}^{\dagger} e^{-i\delta t}-\hat{a} \,e^{i\delta t} ) \nonumber,
\end{eqnarray}
%\begin{center}
\begin{figure}
\hspace{-3.5cm}
\includegraphics[width=11 cm]{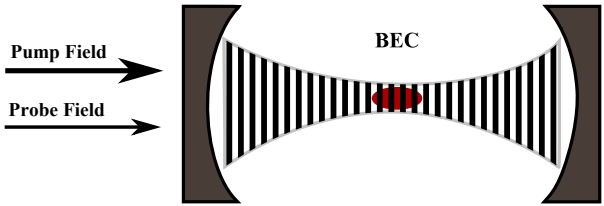}
\caption{Sketch of a BEC-cavity system. A cigar-shaped BEC-cavity system is coupled  to the field of an optical ultrahigh-finesse Fabry-Perot cavity. Two external pump and weak probe lasers  are applied along the cavity axis. The axis of the cigar-shaped BEC and the cavity axis are parallel, so the system is considered one-dimensional.}
\label{figure 1}
\end{figure}
%\end{center}
\noindent where $\Delta=\omega_c-\omega_{pu}$ and $\omega_c$ denotes the cavity oscillation frequency. $V_{ext}({x})$ and $\hat{H}_{A-A}$ are external trapping potential and atom-atom interaction Hamiltonian, respectively. The atomic field operator for the creation of an atom in the ground state with mass $m$, at position $x$, is given by $\hat{\Psi}^{\dagger}({x})$.  Also, $k=\frac{2\pi}{\lambda}$ and $\cos(kx)$  are the wave vector and the cavity mode function, respectively. $U_{0}=g_{0}^{2}/{\Delta_{a}}$  is the maximum light shift which an atom experiences in the cavity mode with the atom-photon coupling constant $g_{0}$. The pump laser frequency is detuned from the atomic $D_{2}$ line transition frequency $\omega_{a}$  by  $\Delta_{a}=\omega_{pu}-\omega_{a} $. Moreover, $\hat{a}^{\dagger}$ and $\hat{a}$ denote the creation and annihilation operators of the cavity mode with the canonical relation, $[\hat{a},\hat{a}^{\dagger}]=1$. In the last term of the Hamiltonian, $\delta=\omega_{pr}-\omega_{pu}$  is the pump-probe detuning.  Using the discrete-mode approximation (DMA)\cite{Zhang,Kozuma}, we can expand $ \hat{\Psi}^{\dagger}({x})$
 into the two spatial modes, $\varphi_{0}$ and  $\varphi_{2}$,
     \begin{equation}
 \hat{\Psi}({x})= \varphi_{0} \hat{c}_{0}+ \varphi_{2}\hat{c}_{2}.
\end{equation}
Here, $\varphi_{0}=1$ and $\varphi_{2}=\sqrt{2} \cos(2kx)$. The Hamiltonian describing the BEC-cavity system for weakly interacting atoms in a shallow external trapping potential, after applying the Bogoliubov approximation and in a rotating frame reads~\cite{Brennecke}:
\begin{eqnarray}
\hat{H}  &=& 4 \hbar\ \omega_{rec} \hat{c}_{2}^{\dagger} \hat{c}_{2} + \hbar \Delta^{\prime} \hat{a}^{\dagger} \hat{a} +\hbar g (\hat{c}_{2}^{\dagger} +\hat{c}_{2} ) \hat{a}^{\dagger} \hat{a} \nonumber\\
&+&i \hbar E_{pu} (\hat{a}^{\dagger}- \hat{a}) + i \hbar E_{pr} ( \hat{a}^{\dagger} e^{-i\delta t}-\hat{a} e^{i\delta t}),
\end{eqnarray}
\noindent where $\hat{c}_{2}^{\dagger}$ and $\hat{c}_{2}$ are the creation and annihilation operators of the Bogoliubov mode with $ \Delta^{\prime}=\Delta+\frac{NU_{0}}{2} $.  The $\varphi_{2}$ is playing the role of a quantum-mechanical oscillator where  $\omega_{m}=4 \omega_{rec}$ denotes its oscillation frequency ($\omega_{rec}$ is the recoil frequency). The parameter $g=\frac{U_{0} \sqrt{N}}{2 \sqrt{2}}$ stands for the effective coupling strength between the cavity and the BEC. The first and the second last terms in the Hamiltonian, describe the classical pump and probe light inputs, where amplitude of the pump (probe) field $E_{pu}$ ($E_{pr}$) is a function of the laser power $P_{pu}$ ($P_{pr}$), according to
\begin{equation}
|E_{pu}|= \sqrt{\frac{2 \kappa P_{pu}}{\hbar \omega_{pu}}},
\end{equation}

\begin{equation}
|E_{pr}|= \sqrt{\frac{2 \kappa P_{pr}}{\hbar \omega_{pr}}},
\end{equation}
\noindent where $\kappa$, is the decay rate of the cavity photons.
According to the Heisenberg equation of motion, we have
\begin{equation}\label{eqa}
\frac{d}{dt} \hat{a}({t})=\\
 (- i \Delta^{\prime} - \frac{\kappa}{2})\hat{a}({t}) - i g \sqrt{2} \hat{X}({t}) \hat{a}({t})  + E_{pu}+ E_{pr} e^{-i\delta t},
\end{equation}
\begin{equation}\label{eqX}
\frac{d^{2}}{dt^{2}} \hat{X}({t})+ \gamma_{m}\frac{d}{dt} \hat{X}({t}) + \omega^{2}_{m}\hat{X}({t})= -\omega_{m} g \sqrt{2}\ \hat{a}^{\dagger}({t})\ \hat{a}({t}),
\end{equation}
\noindent where $\gamma_{m}$ and $\hat{X}=\frac{\hat{c}_{2}^{\dagger} +\hat{c}_{2}}{\sqrt{2}}$ are the damping rate of the atomic excited state and the position operator of the harmonic oscillator, respectively.
By assuming $ \hat{a}({t})=\bar{a}+\delta\hat{a}({t})$ and $ \hat{X}({t})=\bar{x}+\delta\hat{X}({t})$, where $\bar{a}$ and $\bar{x}$ are the interactivity field and mechanical displacement, respectively, the self-consistent steady state solutions are given by
\begin{equation}
\bar{a} = \frac{E_{pu}}{ i\Delta_{c} + \frac{\kappa}{2}},
\end{equation}
\begin{equation}
\bar{x} = \frac{- g \sqrt{2} n}{\omega_{m}},
\end{equation}
\noindent where $n=|\bar{a}|^{2}$ and $\Delta_{c}=\Delta^{\prime}+g \bar{x} \sqrt{2}$.  Now, we insert the ansatz into the equations~(\ref{eqa}) and~(\ref{eqX}) and taking into account them to first order in $\delta \hat{a},\delta \hat{X}$ and $ \delta \hat{a}^{\dagger}$, we obtain
  \begin{eqnarray}
\frac{d}{dt} \delta \hat{a}({t})&=& (- i \Delta_{c} - \frac{\kappa}{2})\delta\hat{a}({t}) \\ \nonumber &-& i g \sqrt{2}\   \delta\hat{X}({t})\ \bar{a}  + E_{pr} e^{-i \delta t},
\end{eqnarray}
\begin{eqnarray}
 \frac{d^{2}}{dt^{2}} \delta\hat{X}(t)&+& \gamma_{m}\frac{d}{dt} \delta\hat{X}({t}) + \omega^{2}_{m}\delta\hat{X}({t})=\\\nonumber &-&\omega_{m} g \sqrt{2}~  \ (\bar{a}^*\delta\hat{a}({t})+\bar{a}~\delta\hat{a}^{\dagger}({t})).
\end{eqnarray}
To solve these coupled equations, we make the ansatz as follows:
\begin{equation}
\delta a(t)= A^{-} e^{-i \delta t} +A^{+} e^{i \delta t},
\end{equation}
\begin{equation}
\delta a^{*}(t)=(A^{+})^{*} e^{-i \delta t} +(A^{-})^{*} e^{i \delta t},
\end{equation}
\begin{equation}
\delta X(t) =X  e^{-i \delta t} +X^{*} e^{i \delta t}.
\end{equation}
\begin{figure*}
\centering
\includegraphics[width=15 cm]{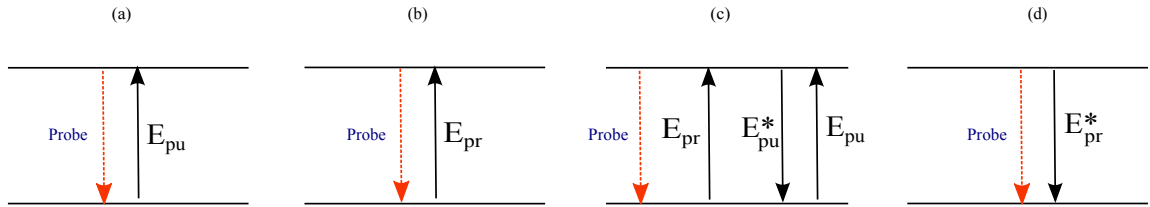}
\caption{Interpretation of the
different contributions to the probe field susceptibility
in terms of transition pathways. (a) represents the scattering of the pump field into the probe field mode. (b) and (c) are the direct linear and nonlinear responses of the BEC-cavity system to the probe fields, respectively.
(d) shows a counter-rotating term, leading to the phase conjugated probe field.}
\label{fig:2}       % Give a unique label
\end{figure*}
Note that, the terms $\bar{a}$, $A^-$ and $A^+$ have simple physical interpretation which is shown in figure 2. According to the equation (8), the first one is corresponding to the $E_{pu}$ and explains the scattering of the pump field into the probe field mode (figure 2-a). The pump field excites the electron from the ground state to the upper level and then it contributes in the probe response. The second one, coefficient of the $e^{-i \delta t}$, is related to the direct response of the BEC-cavity system to the probe fields (figure 2-b and figure 2-c).  The third one represents the counter-rotating contribution, leading to the phase conjugated probe field (figure 2-d). The probe field transmission  depends only on $A^{-}$. So, the three equations of interest are
  \begin{equation}
 A^{-} (- i \delta +i \Delta_{c}+ \frac{\kappa}{2})= E_{pr}- i g~ \bar{a} \sqrt{2}X,
\end{equation}
 \begin{equation}
 (A^{+})^ {*} (- i \delta -i \Delta_{c}+ \frac{\kappa}{2})=  i g~ \bar{a}^* \sqrt{2}X,
\end{equation}
\begin{equation}
 X (\omega^{2}_{m}- i \delta \gamma_{m} - \delta^{2})=  - \omega_{m} g  \sqrt{2}(\bar{a}^*~A^{-}+\bar{a}~(A^{+})^{*}).
\end{equation}
 Solving  Eqs. (15), (16) and (17), we obtain:
 \begin{eqnarray}
 A^{-}&=& \frac{E_{pr}}{\frac{\kappa}{2}- 2 f({\delta})\Delta_{c}+ i(\Delta_{c}-\delta)} \\
\nonumber &+& \frac{i f({\delta})E_{pr}}{\frac{\kappa}{2}- 2 f({\delta})\Delta_{c} +i(\Delta_{c}-\delta)},
 \end{eqnarray}
 with
 \begin{equation}
 f({\delta}) =  \frac{2\omega_{m} g^{2} n \chi({\delta})}{{\frac{\kappa}{2} -i(\delta+\Delta_{c})}},
 \end{equation}
 and
 \begin{equation}
 \chi({\delta})=  \frac{1}{\omega^{2}_{m}- \delta^{2}- i \delta \gamma_{m}}.
 \end{equation}
 According to the equation (18), the first term represents the linear probe response (figure 2-b) while the second term, corresponding to $E_{pr}E_{pu}^2$, explains the cross-Kerr nonlinearity (figure 2-c).
 Using the input-output relations,~\cite{Weis,Cai}
 \begin{equation}
 t=  1- \sqrt{\frac{\kappa}{2}} \frac{A^{-}}{E_{pr}},
 \end{equation}
 the transmission of the probe beam, is given by
\begin{equation}
 t= 1- \frac{(1+i f({\delta}))}{\frac{\kappa}{2}+ 2\delta+i( \Delta_{c}-\delta)}(\frac{\kappa}{2}).
 \end{equation}
In the resolved-sideband regime~\cite{Agarwal}, i.e., $A^+\simeq 0$, one obtains:
 \begin{equation}
 A^{-}= \frac{1}{\frac{\kappa}{2}- i (\delta-\Delta_{c} ) + \frac{g^{2} n}{ \frac{\gamma_{m}}{2}-i (\delta-\Delta_{c})}} \ E_{pr}.
 \end{equation}
We are interested in investigating the group velocity of probe field propagation through the BEC-cavity system. In a dispersive medium, the various frequency components of the light pulse experience different refractive indices.  Therefore, the group velocity of the pulse can change when it enters the dispersive medium. It is well known that the group velocity of light pulse in an anomalous dispersive medium can exceed the light speed in vacuum and even can become negative, leading to the superluminal light propagation. Such propagation does not violate the Einstein's principle of special relativity~\cite{Gaut}.

\section{Results and discussion}
We are going to study the group velocity of a light pulse traveling in a pump-probe BEC-cavity system. Various parameters, i.e., intensity and frequency of the applied fields, damping rate of the atomic excited state  and the decay rate of the cavity photons can be used to control the group velocity in a pump-probe BEC-cavity system. We use the transmission as the indicators to study the group velocity of probe field. According to the Kramers-Kronig relations each peak (dip) in transmission spectrum corresponds to a normal (anomalous) dispersive region on the curve of the refractive index versus frequency, leading to the subluminal (superluminal) light propagation~\cite{Agarwal2, Centini, Li, Jafari}.

Firstly, we investigate the effect of pump field on the transmission of probe field. It is well known that the coupling between the BEC and cavity is strongest at $\Delta_c=\pm \omega_m$~\cite{Agarwal}. Our numerical results are obtained for $\Delta_c=\omega_m$.  In figure 3, we plot the transmission of probe field versus the pump-probe detuning for different values of the pump Rabi frequency. Used parameters are $N=1.2 \times 10^5$, $g_0=2 \pi\times 10.9 MHz$, $\Delta_a=2\pi \times 32 GHz$, $\gamma_m=30\pi KHz$, $\kappa=2\pi\times 1.3 MHz$, $\omega_{rec}=2\pi\times 3.8 KHz$ \cite{Brennecke}, $E_{pu}=.85 MHz$ (solid) and $1.4 MHz$ (dashed). An investigation on the transmission of the probe beam shows that a dip appears around the zero pump-probe detuning and the probe beam propagates superluminally through the cavity. By increasing the intensity of pump field, the gain-doublet response of the medium to the probe field is generated due to the cross-Kerr nonlinearity. Then, according to the Kramers-Kronig relations, the gain assisted superluminal light propagation is established without any attenuation of the probe field~\cite{Wang}.
\begin{figure}
\includegraphics[width=7.0cm]{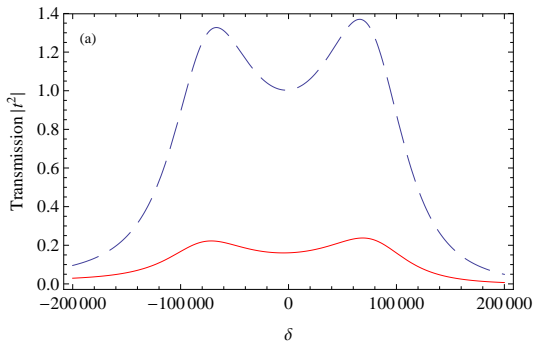}
 \caption{The transmission  of the probe field versus the probe-pump detuning for different values of the pump Rabi frequency. Used parameters are $N=1.2 \times 10^5$, $g_0=2 \pi\times 10.9 MHz$, $\Delta_a=2\pi \times 32 GHz$, $\gamma_m=30\pi KHz$, $\kappa=2\pi\times 1.3 MHz$, $\omega_{rec}=2\pi\times 3.8 KHz$\cite{Brennecke}, $E_{pu}=.85 MHz$ (solid) and $1.4 MHz$ (dashed). }\label{fig3}
\end{figure}
The damping rate of the atomic excited state is another parameter which can affect the optical properties of the BEC-cavity system. Figure 4 illustrates the effect of this parameter on the transmission of the probe field for $E_{pu}=1.4 MHz$, $\gamma_m=30\pi KHz$ (solid) and $90 \pi KHz$ (dashed). Other parameters are same as in figure 3. It can be seen that by increasing the damping rate of the atomic excited state, the dip around $\delta=0$ switches to a peak and the absorption-free superluminal converts to the transparent subluminal light propagation. Therefore, the damping rate of the atomic excited state plays an important role in establishing the gain-assisted superluminal light propagation.
\begin{figure}
\includegraphics[width=7.0cm]{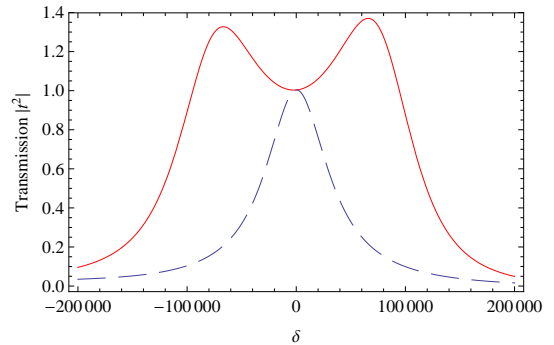}
\caption{The transmission of the probe field versus the probe-pump detuning for $E_{pu}=1.4 MHz$, $\gamma_m=30\pi KHz$ (solid) and $90 \pi KHz$ (dashed). Other parameters are same as in figure 3.}\label{fig4}
\end{figure}
\begin{figure}
\includegraphics[width=7.0cm]{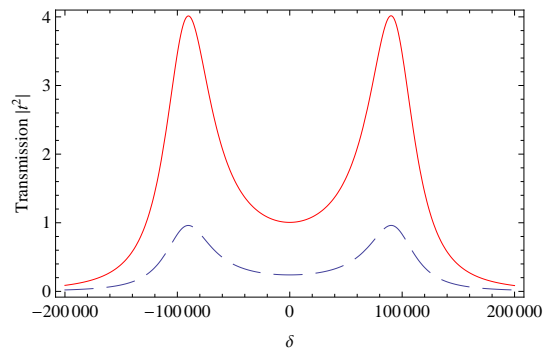}
\caption{The transmission of the probe field versus the probe-pump detuning for different values of the cavity damping, i.e., $\kappa=2\pi \times 24 MHz$ (solid) and $\kappa=2\pi \times 50 MHz$ (dashed). Other used parameters are $E_{pu}=9 MHz$, $\gamma_m=16\pi KHz$, $\Delta_a=2\pi \times 2.6 GHz$.}\label{fig5}
\end{figure}
\begin{figure}
\includegraphics[width=7.5cm]{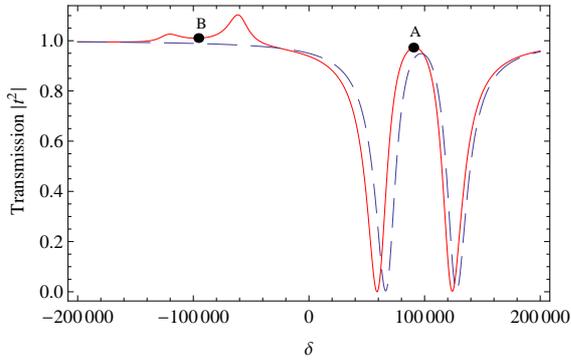}
\caption{The transmission is plotted versus the pump-probe detuning (solid). Used parameters are $\Delta_a=17.4 GHz$, $\gamma_m=2\pi\times 400 Hz$, $\kappa=45 KHz$. Dashed curve shows the transmission of the probe field in the resolved-sideband limit. Points (A) and (B) show the transparent subluminal peak and superluminal dip, respectively.}\label{fig6}
\end{figure}
We are going to study the effect of the decay rate of the cavity photons on the transmission of the probe field. In figure 5, the transmission of the probe field is plotted versus the pump-probe detuning for different values of the decay rate of the cavity photons, i.e., $\kappa=2\pi \times 24 MHz$ (solid) and $\kappa=2\pi \times 50 MHz$ (dashed). Other used parameters are $E_{pu}=9 MHz$, $\gamma_m=16\pi KHz$, $\Delta_a=2\pi \times 2.6 GHz$. It is clear that the decay rate of the cavity photons has a destructive role in establishing the gain-assisted superluminal light propagation.

It is worth noting that, for special set of physical parameters \cite{Frazer}, the transparent subluminal and superluminal light propagation can be realized for different values of the pump field frequency. Such situation is shown in figure 6, in which the transmission is plotted versus the pump-probe detuning (solid). Used parameters are $\Delta_a=17.4 GHz$, $\gamma_m=2\pi\times 400 Hz$ and $\kappa=45 KHz$. It can be seen that the transparent subluminal peak (A) and superluminal dip (B) simultaneously appear in the transmission of probe field around $\delta=\omega_m$ and $\delta=-\omega_m$, respectively. Dashed curve shows the transmission in the resolved-sideband limit, in which the gain doublet due to the effect of the phase conjugation of probe field on the cross-Kerr nonlinearity has been removed. The pump-laser field prepares an optomechanically induced transparent window at $\delta\approx\omega_{m}$ as expected from the analogy with EIT. Two side dips are located at $\delta=\omega_{m}\pm g \sqrt{n}$ in resolved-sideband regime. For $\gamma_{m}=0$, the value at the peak is exactly one and in the dips strict zero is obtained for transmission of the probe field. Note that the role of $\gamma_m$ is similar to that of the decay rate of ground states coherence in the EIT systems~\cite{Agarwal}.
\section{Conclusion}
We have investigated the propagation of a probe field through a cavity optomechanical system with a pump-probe BEC. By calculating the transmission properties of the system, it was shown that the group velocity of the probe pulse field can be controlled by characteristics of the pump laser field. The effect of the decay rate of the cavity photons and damping rate of the atomic excited state on the group velocity was also studied and it was demonstrated that for special experimental parameters, the gain-assisted superluminal light propagation due to the cross-Kerr nonlinearity was established in this system. The gain doublet was removed in the resolved-sideband limit.

\end{document}